**Symmetry induced selective excitation of topological states in Su-Schrieffer–Heeger waveguide arrays**

*Min Tang, * Jiawei Wang, Sreeramulu Valligatla, Christian N. Saggau, Haiyun Dong, Ehsan Saei Ghareh Naz, Sebastian Klembt, Ching Hua Lee, Ronny Thomale, Jeroen van den Brink, Ion Cosma Fulga,* Oliver G. Schmidt, and Libo Ma*

M. Tang, J. Wang, S. Valligatla, C. N. Saggau, H. Dong, E. S. G. Naz, L. Ma
Institute for Integrative Nanosciences
IFW Dresden
Dresden 01069, Germany
E-mail: m.tang@ifw-dresden.de

J. Wang
School of Electronic and Information Engineering
Harbin Institute of Technology (Shenzhen)
Shenzhen 518055, China

S. Klembt
Technische Physik, Wilhelm-Conrad-Röntgen-Research Center for Complex Material Systems, and Würzburg-Dresden Cluster of Excellence ct.qmat
Universität Würzburg
Würzburg D-97074, Germany

C. H. Lee
Department of Physics
National University of Singapore
117551 Singapore

R. Thomale
Institute for Theoretical Physics and Astrophysics,
University of Würzburg
Würzburg D-97074, Germany

J. v. d. Brink, I. C. Fulga
Institute for Theoretical Solid State Physics
IFW Dresden
Dresden 01069, Germany
E-mail: i.c.fulga@ifw-dresden.de

J. v. d. Brink
Institut für Theoretische Physik and Würzburg-Dresden Cluster of Excellence ct.qmat
Technische Universität Dresden
Dresden 01062, Germany

O. G. Schmidt
Research Center for Materials, Architectures and Integration of Nanomembranes (MAIN)
Technische Universität Chemnitz
Chemnitz D-09126, Germany

O. G. Schmidt
Material Systems for Nanoelectronics
Chemnitz University of Technology
Chemnitz D-09111, Germany

O. G. Schmidt
Nanophysics,
Dresden University of Technology
Dresden, D-01069, Germany

**Abstract**

The investigation of topological state transition in carefully designed photonic lattices is of high interest for fundamental research, as well as for applied studies such as manipulating light flow in on-chip photonic systems. Here, we report on topological phase transition between symmetric topological zero modes (TZM) and antisymmetric TZMs in Su-Schrieffer-Heeger (SSH) mirror symmetric waveguides. The transition of TZMs is realized by adjusting the coupling ratio between neighboring waveguide pairs, which is enabled by selective modulation of the refractive index in the waveguide gaps. Bi-directional topological transitions between symmetric and antisymmetric TZMs can be achieved with our proposed switching strategy. Selective excitation of topological edge mode is demonstrated owing to the symmetry characteristics of the TZMs. The flexible manipulation of topological states is promising for on-chip light flow control and may spark further investigations on symmetric/antisymmetric TZM transitions in other photonic topological frameworks.

## 1. Introduction

Topological insulators represent a new class of materials which have a bulk band gap like an ordinary insulator, while exhibiting gapless topological edge or surface states.[1] Topological insulators have been realized in various physical systems that can be characterized by wave equations, including condensed matter systems,[2] acoustic systems,[3] electric circuits,[4] and photonic lattices.[5] Introducing topology into photonics brings unprecedented properties such as topologically protected transport[6] or optical isolation,[7] which has become an intensely investigated field of research. Photonic topological states play a unique role in molding the flow of light,[8] lasing,[9] non-Hermitian photonics,[10] and nonlinear optics.[11] Quasicrystals can also exhibit nontrivial topological properties. The topological nature of these states is utilized to adiabatically pump light across the quasi crystals.[12] The implementation of transitions was achieved by femtosecond laser writing in bulk glass[12b] circuit board printing[12a] which support optical modes in telecom spectral range and surface plasmon polariton modes, respectively. Topological photonic devices with switchable phase transitions are of high interest not only for fundamental studies in controlling lightwave flow with the desired frequency, polarization, and propagating direction, but also for practical applications including optical modulation, filters and on-chip quantum computing.

However, in previous reports, the photonic devices are usually rigid against switching of the topological properties. To investigate topological phase transition, some pioneering works have been reported.[13] For example, spatially structured laser excitations have been utilized to reconfigure non-Hermitian junctions for the control of topological interface states.[13d] By changing polarization states, distinct topological phases and polariton edge states were obtained in perovskite lattices.[13c] Moreover, topological edge mode switching can also be realized by thermal-induced insulator-to-metal transition.[13a] Despite these advances, it is highly demanded to explore and demonstrate novel switching strategies that can realize a transition between different topological states for selective excitation of nontrivial/trivial states.

Here, we design and fabricate reconfigurable one-dimensional (1D) photonic topological insulators composed of $SiN_x$ waveguide arrays for the demonstration of topological state switching. Drawing on an earlier theoretical proposal,[14] the photonic device is based on a Su-Schrieffer-Heeger (SSH) model[15] where the coupling strengths (denoted $\kappa_1$ and $\kappa_2$) can be varied between adjacent waveguide pairs. As shown in Figure 1a and 1b, the dimerized coupling is realized by alternately filling the inter-waveguide gaps with polymethyl methacrylate (PMMA). The waveguide arrays are designed to be symmetrically distributed at the two sides of a center waveguide [see Figure 1c] which serves as the light input. With this symmetric design, depending on the dimerization ($\kappa_1>\kappa_2$ or $\kappa_1<\kappa_2$), either a symmetric TZMs or an antisymmetric TZMs is localized at the central waveguide. For the plasmonic waveguide array with kinked structure, the nonspreading interface mode can be excited in the realistic excitation of single waveguide input.[16] To enable a phase transition between symmetric and antisymmetric TZMs, the air gaps can be filled by ethanol (EtOH) solution [see Figure 1d], which increases the coupling strength therein and in turn enables an inversion of the coupling strength ratio ($\kappa_1/\kappa_2$). Alternatively, the topological phase transition can also be obtained by modifying the PMMA through electron beam irradiation (EBI) which can change the coupling strength at the PMMA filled gaps [Figure 1d]. It should be noted that the gap width is designed to be different to match the coupling strength difference in Figure 1d. With these methods, bidirectional transitions between the symmetric and antisymmetric TZMs are realized in array A and B. Our work presents the switching of the topological phases, which paves the way for the switching of topological states of light to control the light flow in on-chip integrated topological photonic devices.

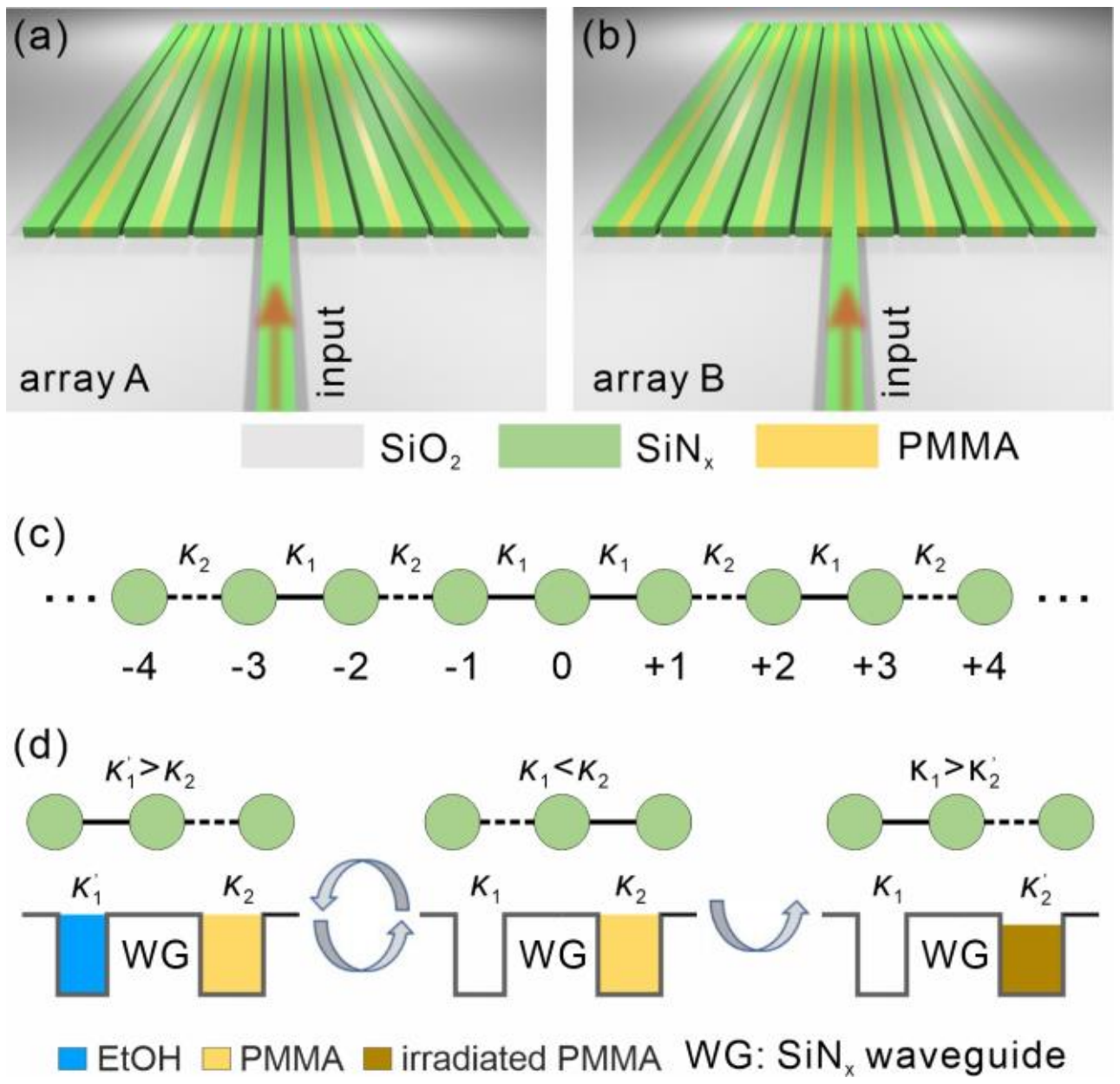

**Figure** 1. Schematic diagram of switchable topological $SiN_x$ waveguide arrays with mirror symmetry at a central waveguide. For a) array A and b) array B, the gaps from the central waveguide to the side waveguides are periodically filled by air-PMMA-air and PMMA-air-PMMA, respectively. c) Schematic of the system with an alternating coupling between neighboring waveguides. In this example, the solid and dashed lines indicate strong and weak coupling, respectively, such that $\kappa_1>\kappa_2$. d) Coupling strength inversion enabled by filling the air gaps with ethanol solution (left panel) or modifying the PMMA (right panel) by electron beam irradiation.

## 2. Results and Discussion

The system can be described based on coupled-mode theory (CMT),[17] which reveals the spatial evolution of optical modes, as well as the eigenmode characteristics. In the present one-dimensional symmetrically-coupled system of $2N+1$ waveguides, $j$ enumerates the waveguides, where the $j^{th}$ waveguide ($j$=0, ±1, …, ±$N$) indicates its relative position to the central waveguide. Each single waveguide has identical geometrical parameters with the same propagation constant $\beta_0$. The coupling strengths from the central to peripheral waveguides are $\kappa_1$ and $\kappa_2$ with periodic alternation. The amplitude $a_j(z)$ of the central and peripheral optical modes evolves along the propagating direction as:

$$i\frac{\partial a_0(z)}{\partial z}+\beta_0 a_0(z)+\kappa_1\big(a_{+1}(z)+a_{-1}(z)\big)=0 \quad (1)$$

$$i\frac{\partial a_j(z)}{\partial z}+\beta_0 a_j(z)+\kappa_1 a_{j+\mathrm{sign}(j)}(z)+\kappa_2 a_{j-\mathrm{sign}(j)}(z)=0$$

(for mod(|$j$|,2) =0) (2)

$$i\frac{\partial a_j(z)}{\partial z}+\beta_0 a_j(z)+\kappa_2 a_{j+\mathrm{sign}(j)}(z)+\kappa_1 a_{j-\mathrm{sign}(j)}(z)=0$$

(for mod(|$j$|,2) =1) (3)

The $a_{j+\mathrm{sign}(j)}(z)$ terms for the edges of the waveguide array in equations (2) and (3) can be discarded because of the open boundary conditions. This treatment is completely analogous to the tight-binding Hamiltonian model in condensed matter physics, where evolving in $z$ direction is an analog to time.[18] Away from the central waveguide, the bulk Hamiltonian takes the same form as the SSH model discussed in a previous report,[19] which implies a similar topological behavior of the TZMs supported by the edge waveguides. For an infinite SSH lattice, the Hamiltonian can be expressed in momentum space as:

$$H(k)=\begin{pmatrix}0 & \rho*(k)\\ \rho(k) & 0\end{pmatrix}\ , \quad (4)$$

where $\rho(k)=\kappa_1+\kappa_2 e^{ik}$. The Hamiltonian exhibits two distinct states depending on the two possible situations of $\kappa_1>\kappa_2$ and $\kappa_1<\kappa_2$, corresponding to trivial and nontrivial topological phases, respectively.[20] The winding number as well as the Zak phase in the Brillouin zone is determined by the $\kappa_1/\kappa_2$ ratio and can be calculated via $\oint dk\langle u_k|\partial_k u_k\rangle$.[21] The Zak phase $Z$ yields $Z$ = 0/π in the case of $\kappa_1>\kappa_2/\kappa_1<\kappa_2$, respectively. A non-zero winding number or Zak phase implies nontrivial topology and thus indicates the existence of topological edge states.

It should be noted that, for both dimerizations ($\kappa_1>\kappa_2$ and $\kappa_1<\kappa_2$), a TZM is present around the central waveguide in our design. This is because the central waveguide represents a defect across which the dimerization pattern changes. The central TZM, however, is either antisymmetric or symmetric to the central waveguide in real space, corresponding to the case of $\kappa_1>\kappa_2$ or $\kappa_1<\kappa_2$. For the case of the symmetric TZM, the maximum optical field is located at the central waveguide, where the optical field distribution can be folded symmetrically ($a_j(z) = a_{-j}(z)$). For the antisymmetric TZM, there exists $a_0(z) = 0$ at the central waveguide and $a_j(z) = -a_{-j}(z)$ for the symmetrically distributed waveguide pairs. As such, symmetric and antisymmetric TZMs with different optical field distributions can be manipulated by changing $\kappa_1/\kappa_2$ in the waveguide arrays.

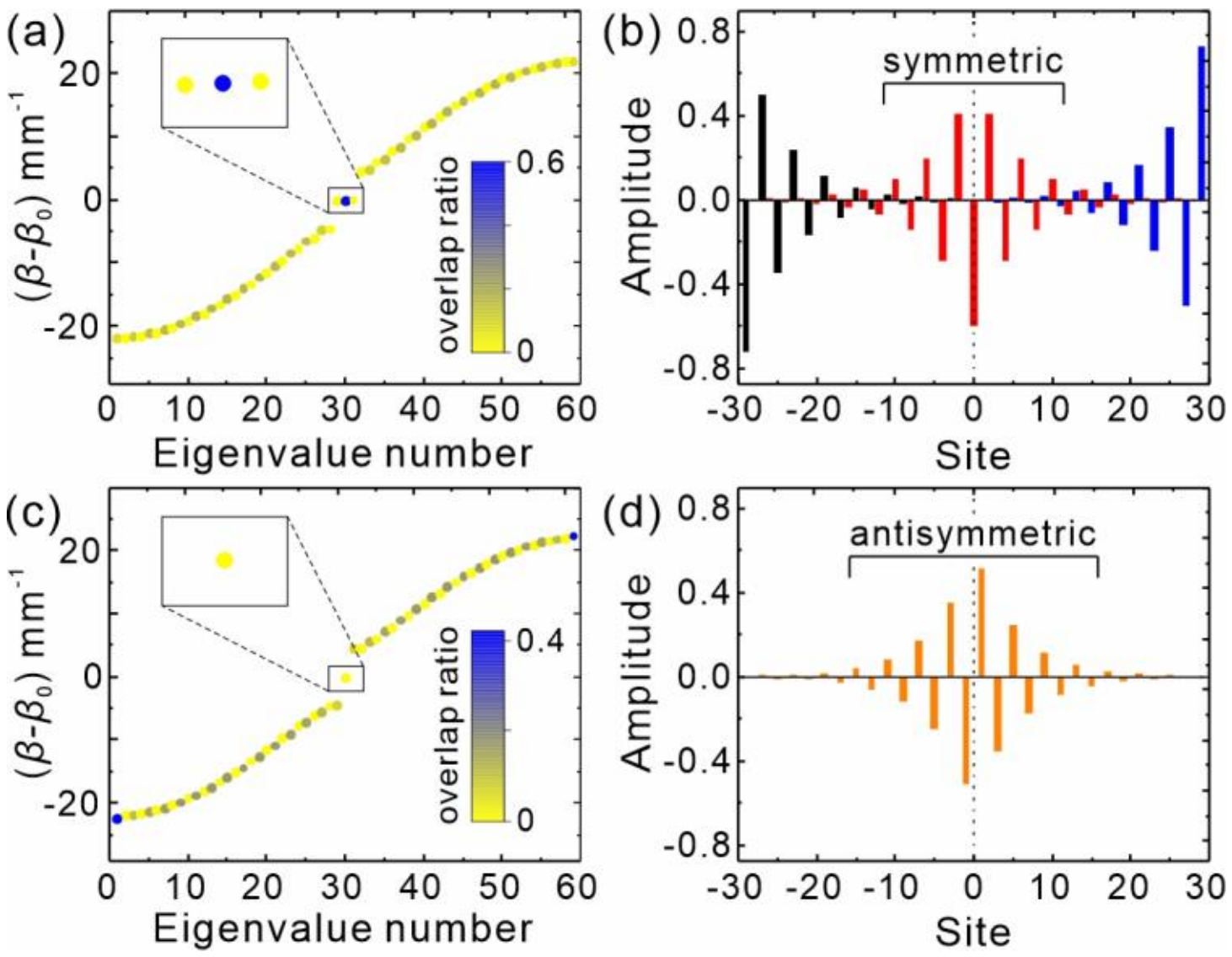

**Figure** 2. Calculated spectrum of the SSH waveguide arrays consisting of 59 waveguides with mirror symmetry to a central waveguide, obtained by solving for the eigenvalues of open boundary Hamiltonian with a) $\kappa_1$=9 mm$^{-1}$, $\kappa_2$=13 mm$^{-1}$, and c) $\kappa_1$=13 mm$^{-1}$, $\kappa_2$=9 mm$^{-1}$. The overlap ratios in a) and c) characterize the overlap between the eigenmode and the central waveguide. b) Real-space distributions of the TZMs at the edge sites (blue and black) and the symmetric TZM at the center waveguide (red, site=0) corresponding to the case of a). d) Real-space distribution of the antisymmetric TZM corresponding to the case of c).

In the fabricated waveguide array with open boundaries, $2N+1=59$ waveguides are designed. Figures 2a and 2c show the results calculated based on the Hamiltonian with open boundary condition, where the bulk band gap opens for both $\kappa_1>\kappa_2$ and $\kappa_1<\kappa_2$, supporting different types of TZM. For the case of $\kappa_1>\kappa_2$, a TZM only occurs at the central part (i.e. waveguides next to the central one) of the waveguide array with an antisymmetric distribution, while a symmetric TZM is present at the central waveguide together with two additional TZMs located at the two sides of the waveguide array for the case of $\kappa_1<\kappa_2$, as shown in Figure 2b and 2d. In the design of input through the central waveguide, the symmetry of the TZM at the central waveguide determines the excitation efficiency of the corresponding states (i.e. the symmetric and antisymmetric TZMs) in the waveguide array. The overlap ratio between the central waveguide and the eigenstate was calculated to be 0.59 for the symmetric TZM as shown in Figure 2a, while the antisymmetric TZM has a zero overlap, as shown in Figure 2c. In other words, the TZM can be efficiently excited via the central waveguide when $\kappa_1<\kappa_2$, while only trivial bulk states can be excited when $\kappa_1>\kappa_2$.

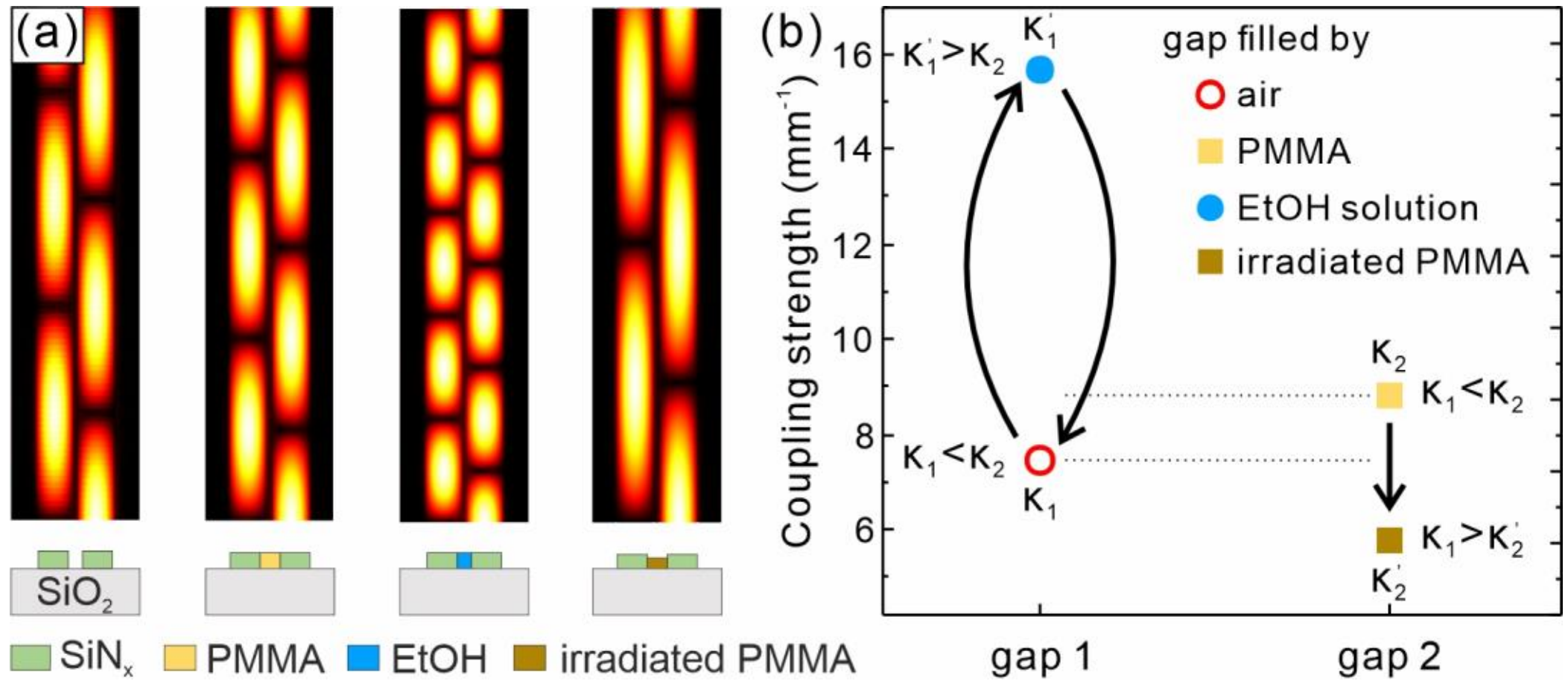

**Figure** 3. a) Simulated optical field ($|H_z|$) distributions under the coupling of fundamental modes propagating along a pair of neighboring waveguides. The gaps of the waveguides are filled by air, PMMA, EtOH solution, and electron-irradiation-modified PMMA. b) The corresponding coupling strength between the waveguides for the four situations.

Based on the above theoretical analysis, we fabricated $SiN_x$ waveguide arrays for the demonstration of topological phase transition by changing the $\kappa_1/\kappa_2$ ratio. Each element of the waveguide array, including the center one, is a $SiN_x$ waveguide with a refractive index $n$=2.02 (@836 nm). The width, height, and length of the $SiN_x$ waveguides are 1.5 μm, 200 nm, and 500 μm, respectively. The waveguide arrays are prepared on a Si substrate spaced by a 2 μm thick $SiO_2$ layer. The $SiO_2$ layer has a refractive index of 1.45, which prevents optical leakage to the substrate. The coupling behavior in a pair of waveguides was examined by the beam propagating method (Figure 3) in a 3D configuration (Comsol Multiphysics 5.3). Gap-1 and 2 in Figure 1 are designed as 120 and 220 nm in width which are filled by air and PMMA, respectively. The inversion of the coupling ratio $\kappa_1/\kappa_2$ is realized by either increasing the coupling strength $\kappa_1$ at the hollow gap-1 or decreasing $\kappa_2$ at gap-2. The increase of $\kappa_1$ can be obtained by filling the hollow gap-1by EtOH solution (refractive index of 1.36), by which the refractive index of gap-1 can significantly increase while that of gap-2 is constant. The decrease of $\kappa_2$ at gap-2 can be achieved by irradiating the filled PMMA using an electron beam (5 keV, 0.02 μC/cm$^2$, 3 min 25 s) in a scanning electron microscope (DSM982, Carl Zeiss Microscopy GmbH) while the $\kappa_1$ at gap-1 is intact during this process. The high-energy electrons can break chemical bonds and significantly change the chemical composition of the filled PMMA.[22] This process generates several effects on the PMMA, including decreasing the refractive index in the near-infrared range,[23] thinning the membrane,[24] and introducing roughness at its surface,[22] all of which will reduce the coupling strength at gap-2. In the simulation, a reduction of 0.1 of the refractive index and 50 nm of the PMMA thickness is considered. The mode distributions of four coupling situations are presented in Figure 3a, corresponding to a gap filled by air, PMMA, EtOH solution, and irradiated PMMA, respectively. The corresponding coupling strength variations from $\kappa$ to $\kappa$' are presented in Figure 3b, showing the capability of inverting the coupling ratio $\kappa_1/\kappa_2$.

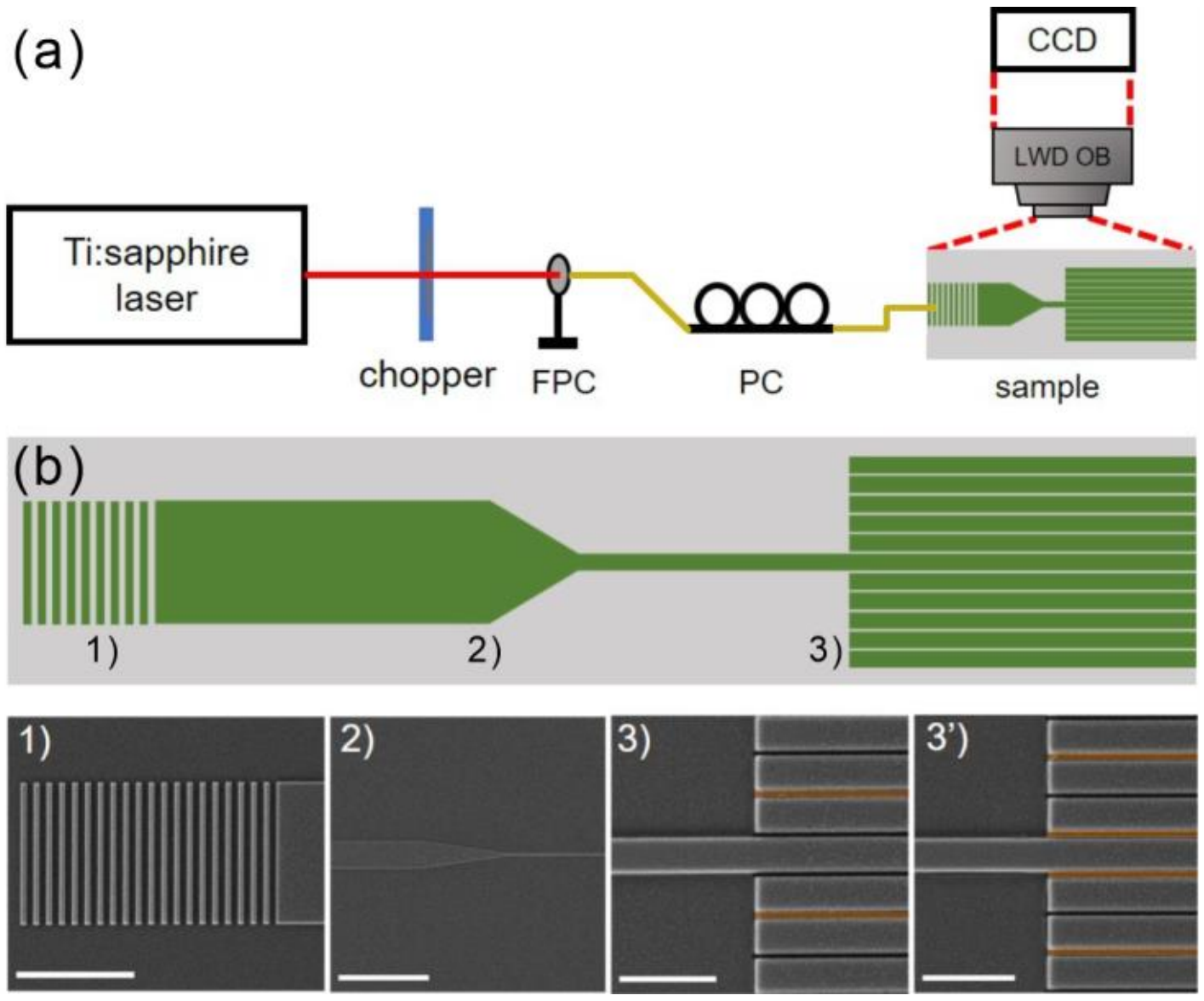

**Figure** 4. a) Schematic diagram of the optical characterization setup. FPC, fiber port coupler; PC, polarization controller; LWD OB, long-working-distance objective lens. The red solid line represents the optic path in free space. The yellow lines represent light guided

through an optical fiber. b) Schematic diagram and SEM images corresponding to different parts of the waveguide array consisting of 1) grating coupler, 2) waveguide mode converter, 3) waveguide array A and 3') waveguide array B. The corresponding scale bars are 10, 40, 4, and 4 μm, respectively. The yellow regions in 3) and 3') indicate the gaps filled by PMMA.

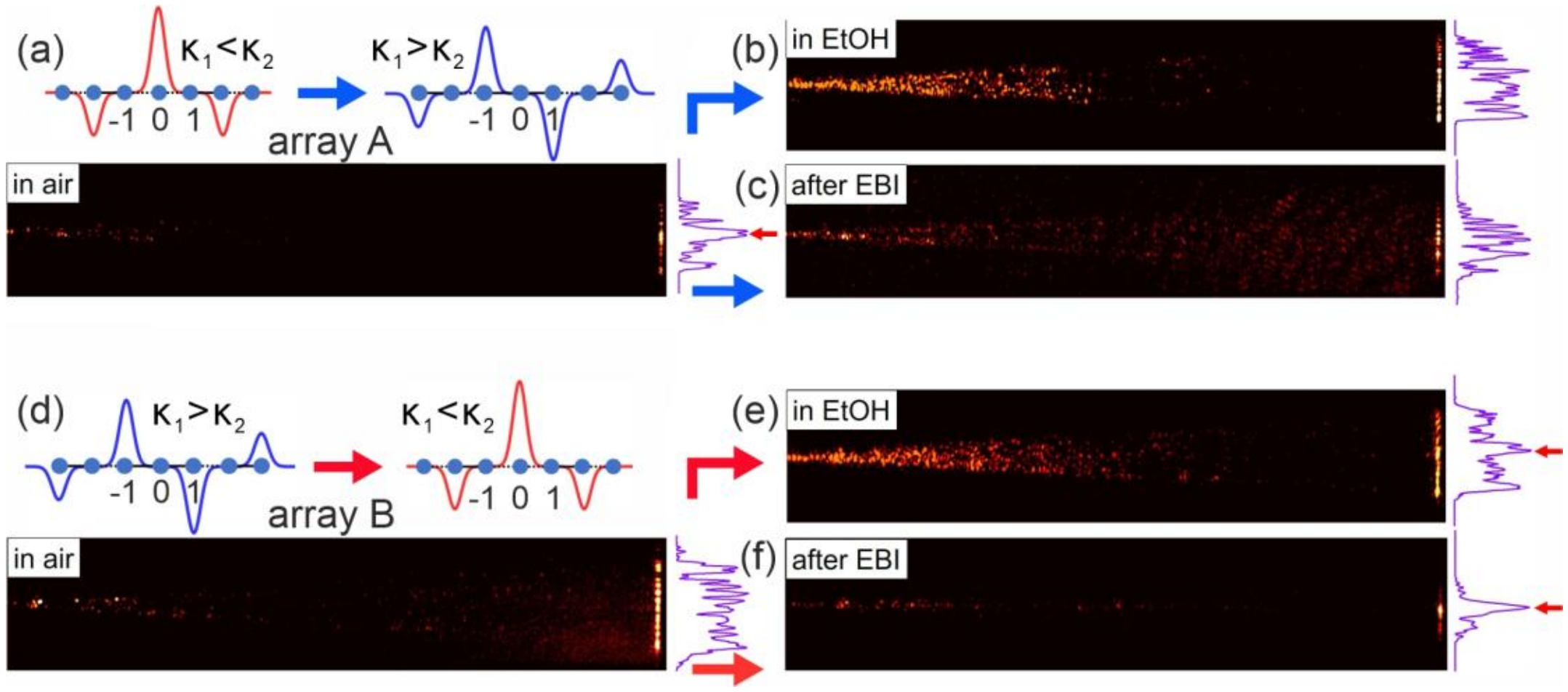

**Figure** 5. Measurement results of the optical transport characteristics in the waveguide array A and B. Panels a), b), and c) show a symmetric to antisymmetric transition of the TZM in waveguide array A, which can be seen from mode field distributions along the waveguides and corresponding field intensity distributions at the end facets of waveguide array. Panels d), e), and f) show the antisymmetric to symmetric transition of the TZM with corresponding field and intensity distributions in the waveguide array B. For a) and d), the sample is in air, while for b) and e), the samples are filled with 90% EtOH solution. For c) and f), the sample is measured after EBI. The height of the image is exactly the same as the waveguide array region while the width of the images is slightly larger than the waveguide length to present the scattering field at the end facets. The schematic only shows the central waveguide site and its nearest sites.

Based on the analysis in Figure 2, manipulating the coupling ratio in the SSH lattice can change the symmetry and distribution of the TZM at the central waveguide, which results in its selective excitation. When the system supports a symmetric TZM ($\kappa_1<\kappa_2$), it can be effectively excited through the central waveguide, as it has a maximum field distribution therein. For the case of an antisymmetric TZM, only bulk states (rather than topological states within the bandgap) can be excited through the central waveguide, as it has a zero distribution of the TZM at the central waveguide. In the experimental verification, the field distribution at the end of the waveguide array is examined to distinguish bulk and in-gap modes excited through the central waveguide. In the air background, the design of array-A and array-B satisfy $\kappa_1<\kappa_2$ and $\kappa_1>\kappa_2$, respectively. Hence, array-A/array-B supports a symmetric/antisymmetric TZM, which has a large/zero overlap with the initial state. The bulk modes can also be excited when $\kappa_1<\kappa_2$ due to the fact that the eigenstate was not used as the initial state to excite the system. Nevertheless, most energy will be transferred into the symmetric TZM and a center peak emerges when being excited through the center waveguide. When $\kappa_1>\kappa_2$, only bulk modes can be excited and a broadened scattering light at the end facets of the waveguide array is observed. In the deposition and etching process, impurity and the rough edge can be introduced to the waveguides. In Figure 5b and 5e, the samples were filled by the EtOH solution, which causes a different distribution of the refractive index in the vertical direction. Compared to the case of air, the refractive index of EtOH solution is closer with the substrate, so the light has a larger probability to couple to the upward loss channel as the refractive contrast of EtOH/$SiN_x$ is smaller than Air/$SiN_x$. In Figure 5c, the irradiation of the electron beam can cause an increase of surface defects and also lead to an increased upward leakage. Topological phase transitions were observed in array A by either filling with 90% EtOH solution or EBI. These processes enable a transition from a symmetric TZM, which can be excited through the central waveguide [see Figure 5a], to an antisymmetric TZM where only trivial bulk modes can be excited through the central waveguide [Figure 5b and 5c]. As shown in Figure 5a, a localized mode was observed at the central waveguide of array A, which is designed to have $\kappa_1<\kappa_2$. After filling with 90% EtOH solution or EBI, the coupling strength is changed to be $\kappa_1>\kappa_2$, where the system supports an antisymmetric TZM. As such, the previously observed TZM disappeared, while only bulk modes can be excited through the central waveguide [Figure 5b and 5c]. A reverse transition from antisymmetric to symmetric TZMs is also realized in array B by the same processing methods, as shown in Figure 5d and 5e-f. Waveguide array B was designed with $\kappa_1>\kappa_2$, supporting an antisymmetric TZM which cannot be excited efficiently through the central waveguide. The coupling ratio was inverted to be $\kappa_1<\kappa_2$ to support a symmetric TZM after filling with 90% EtOH solution or EBI processing. As a result, the topological mode can be excited and observed at the central waveguide [Figure 5e and 5f)]. For the case of $\kappa_1>\kappa_2$, a trivial defect state (TDS) which is known as Tamm state can appear relying on the strong-weak coupling configured at the kinked defect waveguide to the neighboring waveguides.[15a, 15d] For the case of $\kappa_1<\kappa_2$ which supports the symmetric TZM, TDS is not supported due to the weak-strong coupling configuration at the defect region. The localized modes observed in Figure 5a, e and f can only be caused by the excitation of the TZM. For the case of $\kappa_1>\kappa_2$ which in principle can support the antisymmetric TZM and TDS simultaneously, while in our measurement only broadening scattered light at the end facets is observed from Figure 5b-d due to the following two reasons: 1) The antisymmetric TZM is not excited due to the zero overlap with the initial state; 2) The TDS is not strongly localized due to the low coupling strength contrast at the defect region.[15a, 15d] The chemical component change of PMMA is not reversible after electron beam irradiation, However, introducing a switchable material such as organic fluorescent dye controlled through UV/visible photoirradiation[25] could be a potential solution for realizing a reversible topological phase transition by controlling the gap-filling material.

## 3. Conclusion

In summary, topological transitions between symmetric and antisymmetric TZMs have been theoretically and experimentally demonstrated in SSH waveguide arrays designed with mirror symmetry. The transition of TZMs is enabled by manipulating the coupling ratio between neighboring waveguide pairs via liquid filling or electron beam irradiation. Due to the symmetry characteristics of the TZM, either the topological mode or trivial bulk modes can be selectively excited through a central waveguide in the SSH arrays. Our work enables the reshaping of TZMs in the center region of 1D systems for the manipulation and switch of topological states. Moreover, this work is promising for extending to other topological systems, such as the higher-dimensional SSH lattices where the antisymmetric/symmetric TZMs can potentially be designed in multiple directions,[26] non-Hermitian topological photonic systems providing bulk-boundary correspondence,[27] and nonlinear topological systems with nonlinearity-induced topological phase transitions.[11a, 28]

## 4. Experimental Section

The waveguide arrays were fabricated by electron beam lithography (Voyager, Raith, GmbH) and reactive-ion etching (Plasma Lab 100, Oxford Instruments PLC), which is followed by a second-round EBL with careful alignment to define the selective filling of PMMA to the waveguide array gaps. In brief, a 200 nm thick $SiN_x$ layer was first deposited on Si/SiO2 wafer by Plasma-enhanced chemical vapor deposition (ICP: 1200 W; SiH4 [5% in He]: 62 sccm; N2: 20 sccm; Ar: 35 sccm; temperature: 275 °C; time: 1500 s; ICPECVD SI 500 D). PMMA layers (ARP 679.03 and ARP 642.04) with different sensitivities to the electron beam were spin-coated on the $SiN_x$ layer to create a photoresist mask with a negative side wall profile by EBL (50kV, area dose 320 μC/cm2). Then, a 50 nm thick Cr layer was deposited on the patterned PMMA followed by a lift-off process to create a Cr hard mask on the $SiN_x$ layer. The samples were then processed in a fluorine chemistry plasma (45 sccm SF6, 600 W ICP power) to selectively etch away the exposed $SiN_x$ parts. The residual Cr layer was stripped off via a commercial ceric ammonium nitrate/nitric acid-base solution. Finally, a PMMA (ARP 679.03) layer was spin-coated on the sample and the second round of EBL was conducted to remove the PMMA from gap-1.

The sample consists of 1) grating coupler, 2) mode converter, and 3) waveguide arrays. The width and grating period of the grating coupler is 12.0 and 1.0 μm, respectively. Considering the phase-matching condition, the injection fiber angle will be adjusted when the sample is in the air/EtOH solution background. In the measurement setup, a wavelength-tunable Ti:sapphire laser at 836 nm is end-fired into the grating coupler assisted by an optical chopper, fiber-port coupler, and polarization controller. The sample is mounted in a Petri dish which is convenient for solution filling. To image light-scattering patterns from the top view of the waveguide arrays, a long-working-distance microscope objective lens (37 mm) is used together with an 8 MP monochrome CCD camera (S805MU2).

## 5. Topological phase transition analysis

In the following, we will show that the transition from a symmetric to an anti-symmetric TZM in a mirror symmetric SSH chain is a topological phase transition, i.e., it cannot occur without a closing and reopening of the bulk gap.

Consider a mirror-symmetric SSH chain which consists of $2N + 1$ sites and contains a defect in the middle (position $x = 0$), as shown in Figure 1c. The Hamiltonian of the system obeys a chiral symmetry $C$, meaning that $\{H, C\} = 0$, as well as a mirror symmetry $M$, such that $[H, M] = 0$. Though our results are gauge-independent, as we will clarify later, for ease of discussion we begin by considering the site basis, writing:

$$H = \begin{pmatrix} \ddots & & & & & & & \\ & 0 & \kappa_1 & & & & & \\ & \kappa_1 & 0 & \kappa_2 & & & & \\ & & \kappa_2 & 0 & \kappa_1 & & & \\ & & & \kappa_1 & 0 & \kappa_1 & & \\ & & & & \kappa_1 & 0 & \kappa_2 & \\ & & & & & \kappa_2 & 0 & \kappa_1 \\ & & & & & & \kappa_1 & 0 \\ & & & & & & & \ddots \end{pmatrix}, \tag{5}$$

$$M = \begin{pmatrix} & & & & & & & \iddots \\ & & & & & & 1 & \\ & & & & & 1 & & \\ & & & & 1 & & & \\ & & & 1 & & & & \\ & & 1 & & & & & \\ & 1 & & & & & & \\ 1 & & & & & & & \\ \iddots & & & & & & & \end{pmatrix}, \tag{6}$$

and

$$C = \mathrm{diag}(-1, +1, -1, +1, \ldots, +1, -1). \tag{7}$$

The mirror symmetry is spinless, meaning that $M^2 = +1$ and its eigenvalues are $\pm 1$. Notice that since there are an odd number $2N + 1$ of sites in the chain, $M$ is an odd-dimensional matrix, such that it has $N + 1$ eigenvalues equal to $+1$ and only $N$ eigenvalues equal to $-1$.

Since $M$ commutes with $H$, it is possible to label all of the Hamiltonian eigenstates with their corresponding mirror eigenvalue, either $+1$ or $-1$. Mathematically, this means that we can use Schur's Lemma to block-diagonalize the Hamiltonian in the eigenbasis of mirror symmetry. Let $U$ be an $(2N+1) \times (2N+1)$ unitary matrix such that $UMU^{\dagger} = \Lambda_M$, with

$$\Lambda_M = \begin{pmatrix} -\mathbf{1}_{N\times N} & 0 \\ 0 & \mathbf{1}_{(N+1)\times(N+1)} \end{pmatrix} \tag{8}$$

where **1** denotes the identity matrix of a given dimension. By Schur's Lemma, it follows that

$$H_2 = UHU^{\dagger} = \begin{pmatrix} H_- & 0 \\ 0 & H_+ \end{pmatrix} \tag{9}$$

where $H_{\pm}$ are two uncoupled Hamiltonian blocks, of dimensions $N$ and $N+1$, respectively, whose individual eigenstates have mirror eigenvalues $\pm 1$. Note that the possibility of block-diagonalizing the Hamiltonian is a direct consequence of the mirror symmetry, and as such it is gauge-independent. For example, replacing $M \rightarrow e^{i\phi}M$, which is also a valid mirror symmetry, would lead to multiplying $\Lambda_M$ by the same phase, but would not change the block-diagonal nature of $H_2$.

From equations (6) and (7), we see that mirror symmetry commutes with chiral symmetry, $[M, C] = 0$. While this can be easily checked in the specific basis given above, we note that the commutation of $M$ and $C$ is also gauge-independent: $[M, C] = 0$ is preserved under gauge transformations. Due to this fact (and using again Schur's Lemma), chiral symmetry also becomes block-diagonal in the mirror eigenbasis, with

$$C_2 = UCU^{\dagger} = \begin{pmatrix} C_- & 0 \\ 0 & C_+ \end{pmatrix} \tag{10}$$

Thus, $\{H, C\} = \{H_2, C_2\} = 0$ implies that each of the Hamiltonian blocks separately respects chiral symmetry, $\{H_{\pm}, C_{\pm}\} = 0$. In effect, each of the blocks behaves as an independent SSH chain within a given mirror sector. Therefore, each of the two blocks can separately host TZMs, and changing the number of TZMs in any of the two blocks necessarily involves a topological phase transition across which the bulk gap must close.

The above derivation proves that a change from a symmetric to an anti-symmetric TZM at the defect can only be achieved via a topological phase transition. For example, the TZM at the defect in Figure 2b of the main text is even under mirror symmetry, $M\psi_0 = +\psi_0$, meaning that it is necessarily an eigenstate of $H_+$. In Figure 2d, however, we have an anti-symmetric state with $M\psi_0 = -\psi_0$, and the TZM is instead an eigenmode of $H_-$. Any deformation of the Hamiltonian which changes the symmetry of this state must separately change the number of defect zero modes of both $H_-$ and $H_+$. Since each one of them is chiral symmetric, this deformation can only be achieved via a bulk gap closing, marking a topological phase transition.

Physically, when the defect TZM has a mirror eigenvalue −1, this means that $\psi_0(x) = -\psi_0(-x)$, such that $\psi_0(x = 0) = 0$. Therefore, the probability density of this mode must vanish on the central site, which explains why it is not excited [see Figure 5b]. In contrast, when the defect TZM has mirror eigenvalue +1, its probability density at $x = 0$ is not constrained to vanish, such that it can be excited [see Figure 5a].

**Acknowledgements**

This work was financially supported by the Würzburg-Dresden Cluster of Excellence on Complexity and Topology in Quantum Matter–ct.qmat (EXC 2147, project-ID 390858490). L.B.M. acknowledges financial support by the German Research Foundation (MA 7968/2-1). J. W. acknowledges the support of the National Natural Science Foundation of China (Grant No. 62105080) and Program for International S&T Cooperation Projects of GuangDong Province (Grant No. 2022A0505030003). O.G.S. acknowledges financial support by the Leibniz Program of the German Research Foundation (SCHM 1298/26-1). The authors thank M. Bauer, L. Raith and R. Engelhard for the technical support; S. Baunack for helpful discussion and SEM characterizations.

Received: ((will be filled in by the editorial staff))

Revised: ((will be filled in by the editorial staff))

Published online: ((will be filled in by the editorial staff))